\documentclass[11pt,twoside,onecolumn]{article}
\usepackage[]{latexsym}
\usepackage[dvips]{epsfig}
\newcommand{\be}{\begin{eqnarray}}
\newcommand{\ee}{\end{eqnarray}}

\title{\bf On boundary terms and conformal transformations in curved
space-times}
\author{R. Casadio\thanks{casadio@bo.infn.it}$\ $
and
A. Gruppuso\thanks{gruppuso@bo.infn.it}
\\
\\
{\em Dipartimento di Fisica, Universit\`a di Bologna and
I.N.F.N., Sezione di Bologna,}\\
{\em via Irnerio 46, 40126, Bologna, Italy}}
\begin{document}
%
%
\maketitle
\begin{abstract}
We intend to clarify the interplay between boundary terms and
conformal transformations in scalar-tensor theories of gravity.
We first consider the action for pure gravity in five dimensions and
show that, on compactifing {\em a la\/} Kaluza-Klein to four
dimensions, one obtains the correct boundary terms in the Jordan
(or String) Frame form of the Brans-Dicke action.
Further, we analyze how the boundary terms change under the conformal
transformations which lead to the Pauli (or Einstein) frame and to
the non-minimally coupled massless scalar field.
In particular, we study the behaviour of the total energy in
asymptotically flat space-times as it results from surface terms
in the Hamiltonian formalism.
\end{abstract}
%
%
\raggedbottom
\setcounter{page}{1}
\section{Introduction}
\setcounter{equation}{0}
\label{intro}
It is known that there is a chain of conformal transformations which
relates higher-dimensional pure gravity compactified to four dimensions
in the so called Jordan (or String) frame (SF) to the Pauli (or Einstein)
frame (EF) with a minimally coupled scalar field (for a review, see
{\em e.g.}, Ref.~\cite{overduin}) and the latter to a non-minimally
coupled scalar field \cite{faraoni} (and Refs. therein).
This issue is however not free of controversy, the main points being
the consistency of the original Kaluza-Klein (KK) compactification and
which ``frame'' is to be taken in four dimensions as that possibly
describing real-world Physics \cite{faraoni}.
\par
In the present paper we focus on another aspect which is usually
``overlooked'' in the literature, namely the role played by boundary
terms in the various forms of the action and the way they change under
such conformal transformations.
For simplicity, we shall consider the case of five-dimensional
pure gravity, for which there is no ``ground state'' problem (Minkowski
space can be easily recovered when there is just one extra dimension).
Our starting point is the Einstein-Hilbert action on the
five-dimensional space-time manifold $\hat{\cal M}$,
\be
\hat S={1\over 16\,\pi\,\hat G}\,\int_{\hat{\cal M}}
d^5x\,\sqrt{-\hat g}\,\hat R
\ ,
\label{S5}
\ee
where $\hat G$ is Newton constant in five dimensions, $\hat g_{AB}$
the five-dimensional metric ($A,B=0,\ldots,4$ and $\mu,\nu=0,\ldots,3$),
\be
\hat g_{AB}=\left[\begin{array}{cc}
\bar g_{\mu\nu}+\kappa^2\,\hat\phi^2\,A_\mu\,A_\nu &
\kappa\,\hat\phi^2\,A_\nu \\
\kappa\,\hat\phi^2\,A_\mu & \hat\phi^2
\end{array}\right]
\ ,
\ee
and $\hat R$ its scalar curvature.
Upon defining $\hat\phi=e^{-\phi}$ and compactifying {\em a la\/}
KK~\footnote{The basic assumptions are that the five-dimensional
space-time has cylindrical symmetry, $\hat{\cal M}={\cal M}\times S^1$,
and all fields do not depend on the fifth dimension.}
the metric and then setting the constant $\kappa=0$ one can
arrive at the Jordan (or String frame - SF) form of the
Brans-Dicke action (without matter fields)
\be
S_{SF}=\int_{\cal M} d^4x\,\sqrt{-\bar g}\,
e^{-\phi}\,\left[{\bar R\over 16\,\pi\,G}
+\omega\,\left(\bar \nabla_\mu\phi\right)\,
\left(\bar \nabla^\mu\phi\right)
\right]
\ .
\label{S_SF}
\ee
In the above, the scalar field $\phi$ is also called the {\em dilaton}
in string theory \cite{gsw}, $\bar R$ is the scalar curvature and
$\bar\nabla$ the covariant derivative for the four-dimensional metric
$\bar g_{\mu\nu}$, and the four-dimensional Newton constant
$G=\hat G/V_S$, where $V_S=2\,\pi$ is the ``coordinate volume'' of
the circle $S^1$.
We shall set $8\,\pi\,G=1$ and $\omega=1/4$ from now on, since
the following considerations hold in general and any choice of these
constants does not affect the analysis of boundary terms.
\par
The conformal transformation
\be
\bar g_{\mu\nu}=\Omega^2\,g_{\mu\nu}
\ ,
\label{T}
\ee
with
\be
\Omega=e^{\phi/2}
\ ,
\label{Tgg}
\ee
yields the Pauli (or Einstein frame - EF) form
of the action
\be
S_{EF}[g_{\mu\nu},\phi]=
{1\over 2}\,\int_{\cal M} d^4x\,\sqrt{-g}\,
\left[R-\left(\nabla_\mu\phi\right)\,\left(\nabla^\mu\phi\right)\right]
\ ,
\label{S_EF}
\ee
in which $\phi$ is minimally-coupled to gravity and has zero conformal
weight.
\par
A non-minimal coupling (parameterized by the constant $\xi$) can
finally be introduced by transforming the metric according to
\be
\tilde g_{\mu\nu}=\Omega^2_\xi\,g_{\mu\nu}
\ ,
\label{Tg}
\ee
where $\Omega^2_\xi\equiv\left(1-\xi\,\tilde\phi^2\right)^{-1}$.
The new scalar field is related to $\phi$ by
\be
{d\tilde\phi\over d\phi}=
{1-\xi\,\tilde\phi^2\over
\left[1-\xi\,\left(1-6\,\xi\right)\,\tilde\phi^2\right]^{1/2}}
\ .
\label{Tphi}
\ee
We note that $\tilde\phi$ cannot be assigned a specific conformal weight
for generic $\xi$.
Explicit solutions of Eq.~(\ref{Tphi}) are rather involved, except for
the conformally coupled case $\xi=1/6$ (see Appendix~\ref{1/6}).
In general the action reads
\be
\tilde S[\tilde g_{\mu\nu},\tilde\phi]
={1\over 2}\,\int_{\cal M} d^4x\,\sqrt{-\tilde g}\,\left[
\left(1-\xi\,\tilde\phi^2\right)\,\tilde R
-\left(\tilde\nabla_\mu\tilde\phi\right)\,
\left(\tilde\nabla^\mu\tilde\phi\right)
\right]
\ ,
\label{tS}
\ee
where $\tilde \nabla_\mu$ and $\tilde R$ represent the covariant
derivative with respect to and the scalar curvature of the metric
$\tilde g_{\mu\nu}$.
\par
In the above manipulations we have always neglected all possible
boundary terms at the border $\partial{\cal M}$ of ${\cal M}$
which arise from the scalar curvature.
In fact, the latter changes under a conformal transformation
(\ref{T}) according to \cite{birrell},
\be
R\to \Omega^{-2}\,\left[R
-2\,(d-1)\,\Omega^{-1}\,\nabla_\mu\nabla^\mu\Omega
-(d-1)\,(d-4)\,\Omega^{-2}\,\nabla_\mu\Omega\,\nabla^\mu\Omega
\right]
\ ,
\ee
which holds in $d$ space-time dimensions and for any $\Omega$.
However, such terms must be handled carefully \cite{gh,regge,wald}
in order to obtain the correct equations of motion both in the
Lagrangian and Hamiltonian formalism.
This is what we shall deal with in the next Sections.
\par
In Section~\ref{Lag} we consider the role of boundary terms in the
Lagrangian formalism and, in Section~\ref{Ham}, in the Hamiltonian
formalism.
The latter case is particularly interesting because the surface terms
can therein be related to the value of the canonical Hamiltonian of
the system, that is, the ADM mass in asymptotically flat space-times.
Particular cases are then shown in Appendix~\ref{large}.
\section{Boundary terms in the Lagrangian formalism}
\setcounter{equation}{0}
\label{Lag}
From the point of view of the Lagrangian formalism, the
Einstein-Hilbert (EH) action,
\be
S_{EH}={1\over 2}\,\int_{\cal M} d^4x\,\sqrt{-g}\,R
\ ,
\ee
is ill-defined if the border of the space-time manifold,
${\partial\cal M}$, is not an empty set.
In fact, the equations of motion in the Lagrangian formalism
are obtained by varying the action with the value of the field
variables held fixed at the boundary, namely
\be
\left.\delta g_{\mu\nu}\right|_{\partial\cal M}=0
\ .
\ee
However, the presence of second derivatives of the metric
inside the scalar curvature $R$ would also require the condition
\be
\left.
\delta\partial_\lambda g_{\mu\nu}\right|_{\partial\cal M}=0
\ ,
\label{dd}
\ee
which can be disposed of by subtracting from the volume part of
the EH action the surface term containing (minus) the (trace of the)
extrinsic curvature $K$ of the boundary \cite{gh,wald},
\be
S_{EH}\to S'_{EH}=S_{EH}+
\int_{\partial\cal M} d^3\zeta\,\sqrt{|\gamma|}\,K
\ ,
\label{bEH}
\ee
where $\gamma_{ij}$ is the three-metric on $\partial{\cal M}$ and
$S'_{EH}$ contains no term linear in second derivatives of the metric.
\par
It is straigtforward to generalize the above prescription to the
five-dimensional action (\ref{S5}),
\be
\hat S\to \hat S'=\hat S+
\int_{\partial\hat{\cal M}} d^4\zeta\,\sqrt{|\gamma^{(4)}|}\,K^{(4)}
\ ,
\label{b5}
\ee
where $\gamma^{(4)}_{\mu\nu}$ is the metric on the four-dimensional
border $\partial\hat{\cal M}$ and $K^{(4)}$ the trace of its extrinsic
curvature.
After the dimensional reduction~\footnote{We assume the border of the
five-dimensional manifold also has cylindrical symmetry,
$\partial\hat{\cal M}=\partial{\cal M}\times S^1$}, from Eq.~(\ref{b5}) one
obtains
\be
S_{SF}'[\bar g_{\mu\nu},\phi]=S_{SF}[\bar g_{\mu\nu},\phi]
+\int_{\partial\cal M} d^3\zeta\,\sqrt{|\bar\gamma|}\,
e^{-\phi}\,\bar K
\ .
\label{S_SF'}
\ee
We shall now show that, as one would expect, this is precisely the
generalization of the prescription (\ref{bEH}) which is required
to have a well-defined Lagrangian form.
\subsection{String Frame and Einstein Frame}
\label{Lse}
Upon changing to the EF according to the transformation (\ref{Tgg})
one obtains
\be
S_{SF}[\bar g_{\mu\nu},\phi]=
S_{EF}[g_{\mu\nu},\phi]
-{3\over 2}\,\int_{\cal M} d^4x\,\sqrt{-g}\,\left(
\nabla_\mu\nabla^\mu\phi\right)
\ .
\label{bSF}
\ee
One now notices that
\be
\sqrt{|\bar\gamma|}\,e^{-\phi}\,\bar K
=\sqrt{|\gamma|}\,K
+{3\over 2}\,\sqrt{|\gamma|}\,n^\mu\,\nabla_\mu\phi
\ ,
\ee
where $n^\mu$ is the (time- or space-like) normal to
$\partial{\cal M}$.
One can therefore conclude that the second term in the right hand
side (r.h.s.) of Eq.~(\ref{bSF}) is exactly cancelled against the
second term above, since
\be
\int_{\cal M} d^4x\,\sqrt{-g}\,\nabla_\mu\nabla^\mu\phi
=\int_{\cal M} d^4x\,\partial_\mu\left(\sqrt{-g}\,\nabla^\mu\phi\right)
=\int_{\partial\cal M} d^3\zeta\,\sqrt{|\gamma|}\,n^\mu\,\nabla_\mu\phi
\ .
\ee
Hence
\be
S'_{SF}[\bar g_{\mu\nu},\phi]
=S_{EF}[g_{\mu\nu},\phi]
+\int_{\partial\cal M} d^3\zeta\,\sqrt{|\gamma|}\,K
\equiv
S'_{EF}[g_{\mu\nu},\phi]
\ .
\label{bSF1}
\ee
In other words, if one takes properly into account the boundary
terms in the action (\ref{S_SF'}), no boundary terms are generated
by the transformation (\ref{Tgg}) except those
which are required to dispose of the (unwanted) condition on
the first derivatives of the metric, such as that in Eq.~(\ref{dd}).
\subsection{Non-minimally coupled scalar field}
\label{Lnmc}
The action for a non-minimally coupled, massless scalar field
$\tilde\phi$ in a space-time ${\cal M}$ with metric
$\tilde g_{\mu\nu}$ is given by Eq.~(\ref{tS}) \cite{birrell}.
On carefully performing the transformations (\ref{Tg}) and
(\ref{Tphi}), one  actually finds that a boundary term is
also generated, so that
\be
\tilde S[\tilde g_{\mu\nu},\tilde\phi]
&=&S_{EF}[g_{\mu\nu},\phi]
-3\,\int_{\cal M} d^4x\,\sqrt{-g}\,\left(
\nabla_\mu\nabla^\mu\ln\Omega_\xi\right)
\nonumber \\
&=&S_{EF}[g_{\mu\nu},\phi]
-3\,\int_{\partial\cal M} d^3\zeta\,\sqrt{|\gamma|}\,
n^\mu\,\nabla_\mu\ln\Omega_\xi
\ .
\label{bS}
\ee
In order to eliminate second derivatives of the metric, one can now employ
the following prescription \cite{acg} instead of (\ref{bEH})
\be
\tilde S[\tilde\phi,\tilde g_{\mu\nu}]\to
\tilde S'[\tilde\phi,\tilde g_{\mu\nu}]
=\tilde S[\tilde\phi,\tilde g_{\mu\nu}]
+\int_{\partial\cal M} d^3\zeta\,\sqrt{|\tilde\gamma|}\,
\left(1-\xi\,\tilde\phi^2\right)\,\tilde K
\ .
\label{pre}
\ee
Since
\be
\sqrt{|\tilde\gamma|}\,
\left(1-\xi\,\tilde\phi^2\right)\,\tilde K
=\sqrt{|\gamma|}\,K
+3\,\sqrt{|\gamma|}\,n^\mu\,\nabla_\mu\ln\Omega_\xi
\ ,
\ee
one again finds that
\be
\tilde S'[\tilde g_{\mu\nu},\tilde \phi]
=S_{EF}[g_{\mu\nu},\phi]
+\int_{\partial\cal M} d^3\zeta\,\sqrt{|\gamma|}\,K
\equiv
S'[g_{\mu\nu},\phi]
\ ,
\label{tSB}
\ee
and concludes that unwanted boundary terms are not generated
by the transformations (\ref{Tg}) and (\ref{Tphi}), in complete
analogy with the case of Section~\ref{Lse}.
\section{Boundary terms in the Hamiltonian formalism}
\setcounter{equation}{0}
\label{Ham}
When one moves on to the Hamiltonian formalism, the handling of
surface terms becomes more subtle because they acquire a dynamical
meaning.
One first performs the ADM decomposition \cite{adm} of the metric,
\be
g_{\mu\nu}=\left[\begin{array}{cc}
-N^2+N_k\,N^k & N_j \\
N_i & \gamma_{ij}
\end{array}\right]
\ ,
\ee
where $N$ is the lapse function and $N^i$ are the shift functions
associated to a given foliation of ${\cal M}$ into spatial
hypersurfaces $\Sigma_t$ whose unit time-like normal is denoted by
\be
t^\mu=\left({1\over N}\ ,\ -{N^i\over N}
\right)
\ .
\ee
The EH Lagrangian can then be written as
\be
L_{EH}={1\over 2}\,\int_{\Sigma_t} d^3x\,\sqrt{-g}\,R
\ ,
\ee
with
\be
\sqrt{-g}\,R&=&
N\,\sqrt{\gamma}\,\left(K_{ij}\,K^{ij}-K^2+R^{(3)}\right)
\nonumber \\
&&-2\,\left[\sqrt{\gamma}\,K\right]_{,t}
+2\,\left[\sqrt{\gamma}\,\left(K\,N^j-\gamma^{ij}\,N_{,i}\right)
\right]_{,j}
\ ,
\label{L}
\ee
where
\be
K_{ij}&=&{1\over 2\,N}\,
\left(N_{i|j}+N_{j|i}-\gamma_{ij,t}\right)
\ ,
\ee
is the extrinsic curvature tensor of the hypersurfaces $\Sigma_t$,
$|$ denotes the covariant derivative with respect to the
three-metric $\gamma_{ij}$ and $R^{(3)}$ is its intrinsic curvature.
Second time derivatives of the three-metric $\gamma_{ij}$ should
again not appear in the action and this is accomplished
by adding a surface term of the form in Eq.~(\ref{bEH}) on
the initial ($\Sigma_{t_1}$) and final ($\Sigma_{t_2}$) slices,
\be
\left[\int_{\Sigma_t} d^3x\,\sqrt{|\gamma|}\,K\right]_{t_1}^{t_2}
=\int_{t_1}^{t_2} dt\,\int_{\Sigma_t} d^3x\,
\left[\sqrt{\gamma}\,K\right]_{,t}
\ ,
\label{Bt}
\ee
which cancels against the second term in the r.h.s. of
Eq.~(\ref{L}) above.
However, one finds that the surface terms at $\partial\Sigma_t$
which arise from $R^{(3)}$ and the third term in Eq.~(\ref{L})
are related to the energy and momentum of the system, as we review
below.
\par
Since we are mainly concerned with symptotically flat space-times,
from now on we assume that the border of $\Sigma_t$ is a two-sphere
and introduce (asymptotically, {\em i.e.}, at $r\to\infty$) spherical
coordinates $(r,\theta,\varphi)$ such that $\partial\Sigma_t\equiv\{r=R\}$.
It is then consistent to impose the following conditions \cite{regge}
\be
\begin{array}{l}
\gamma_{ij}-\eta_{ij}
\sim N-1
\sim N^i
\sim r^{-1}
\\
\\
\gamma_{ij,k}
\sim N_{,k}
\sim N^i_{,k}
\sim r^{-2}
\ .
\end{array}
\label{asy}
\ee
One can thus neglect terms containing the $N^i$'s and,
after adding the term in Eq.~(\ref{Bt}) above, one obtains
the Lagrangian~\footnote{Generalization to the case when
there is a matter source is straightforwardly obtained by
adding the matter Lagrangian, $L^{(0)}_{EH}\to L^{(0)}_{EH}+
L_{M}$.
\label{general}}
\be
L^{(0)}_{EH}&=&{1\over 2}\,\int_{\Sigma_t} d^3x\,
N\,\sqrt{\gamma}\,\left(K_{ij}\,K^{ij}-K^2+R^{(3)}\right)
-\int d\theta\,d\varphi\,
\left[\sqrt{\gamma}\,\gamma^{ri}\,N_{,i}
\right]_{r=R}
\ .
\label{Va}
\ee
Since $L_{EH}^{(0)}$ still contains second (spatial) derivatives,
it does not meet the prescription required by the Lagrangian
formalism and, as a consequence, upon varying the Hamiltonian
($\pi^{ij}\equiv{\delta L_{EF}^{(0)}\over \delta\gamma_{ij,t}}$
are the canonical momenta conjugated to $\gamma_{ij}$)
\be
H^{(0)}_{EH}=
\int_{\Sigma_t} d^3x\,\left(\pi^{ij}\,\gamma_{ij,t}\right)
-L^{(0)}_{EH}
\ ,
\ee
one gets unwanted terms in the Hamiltonian form of the field
equations \cite{regge}.
Such terms can be eliminated by adding a suitable surface term
to the Hamiltonian $H_{EF}^{(0)}$, for which just the asymptotic
form is usually given \cite{dewitt,regge,wald}.
\par
For the sake of simplicity, we assume for the three-metric
the form
\be
\gamma_{ij}\,dx^i\,dx^j
=f^2\,dr^2+h^2\,\left(d\theta^2+\sin^2\theta\,d\varphi^2\right)
\ ,
\label{g3}
\ee
where $f=f(r)$ and $h=h(r)$, and neglect the $N^i$'s in agreement
with the conditions in Eq.~(\ref{asy}) and spherical symmetry.
The three-curvature thus gives rise to a term
\be
N\,\sqrt{\gamma}\,R^{(3)}\supseteq -N\,f\,h^2\,\sin\theta\,
{4\,h_{,rr}\over h\,f^2}
=-\left[4\,N\,\sin\theta\,{h\,h_{,r}\over f}\right]_{,r}+\ldots
\ ,
\ee
which, once added to the second term in the r.h.s. of Eq.~(\ref{Va}),
yields
\be
-\int d(\cos\theta)\,d\varphi\,\left(r^2\,N\,N_{,r}+2\,r\,N^2\right)_{r=R}
=-\int_{\partial\Sigma_t} d\theta\,d\varphi\,\sqrt{\gamma_{R}}\,K_R
\equiv E_{EH}
\ .
\label{E_EH}
\ee
In Eq.~(\ref{E_EH}) we have set
\be
f=N^{-1}\ ,\ \ \ \ h=r
\ ,
\label{N=f}
\ee
and $\gamma_R$ is the three-metric of the surface $r=R$ with $K_R$ the
corresponding extrinsic curvature.
\par
Unwanted terms can now be eliminated from the Hamiltonian by
subtracting from $L_{EH}^{(0)}$ the usual boundary term
\be
L_{EH}'=L_{EH}^{(0)}-E_{EH}
\ .
\ee
from which one recovers the expressions given in \cite{dewitt,regge,wald}
for $R\to\infty$.
\par
It is now interesting to recall that $L'_{EH}$ reduces to $-E_{EH}$
when it is evaluated for a (static) solution of the field equations.
This is not surprising, since any solution satisfies the constraints
corresponding to the reparameterization invariances of General Relativity,
\be
{\cal H}={\cal H}_i=0
\ ,
\ee
where ${\cal H}$ and ${\cal H}_i$ are implicitly defined by
\be
H^{(0)}_{EH}=
\int_{\Sigma_t} d^3x\,\left(N\,{\cal H}+N^i\,{\cal H}_i\right)
\ .
\ee
For a static space-time $\gamma_{ij,t}=0$, thus the above
constraints imply that
$H^{(0)}_{EH}=L^{(0)}_{EH}=0$~\footnote{In general
$L^{(0)}=0$ for a static source (see footnote~\ref{general}).}.
Therefore, the total Hamiltonian becomes
\be
H_{EH}'\equiv H_{EH}^{(0)}+E_{EH}=E_{EH}
\ ,
\label{E_EH1}
\ee
and is proportional to the ADM mass $M$ contained within the sphere
of radius $R$ in the limit $R\to\infty$ (provided one subtracts
a flat space contribution \cite{gh}, see also Appendix~\ref{large})
\par
Under a general conformal transformation (\ref{Tg}) one expects
that the value of the energy as measured locally changes, since
such a transformation represents a (local) rescaling of the units
\cite{quiros} according to the relation
\be
\bar E=\Omega^{-1}\,E
\ ,
\label{a}
\ee
which follows from naive dimensional analysis.
However, the same argument does not apply to the ADM mass,
since $M$ is a global quantity related to the total energy
contained in the system.
In fact, one can consider the geodesic equations at large $r$
in the Newtonian limit and for a static space-time in the
rescaled frame one obtains
\be
{d^2r\over dt^2}&\simeq& {1\over 2}\,\bar g^{rr}\,\partial_r\bar g_{tt}
\nonumber \\
&=&
{1\over 2}\,g^{rr}\,g_{tt,r}
+g^{rr}\,g_{tt}\,\partial_r(\ln\Omega)
\ ,
\label{geo}
\ee
where the first term in the r.h.s. above is the unrescaled frame
contribution and the second term thus represents a correction.
If $\Omega$ contains a term of order $r^{-1}$,
\be
\Omega\sim 1-{\delta M\over 8\,\pi\,r}
\ ,
\label{O_large}
\ee
where $\delta M$ is a constant (not necessarily small),
and $N$ is as in Eq.~(\ref{schw}), then Eq.~(\ref{geo})
yields
\be
{d^2r\over dt^2}\simeq -{\bar M\over 8\,\pi\,r^2}
\ ,
\ee
where $\bar M=M+\delta M$.
However, upon substituting the form (\ref{O_large}) into
Eq.~(\ref{a}), one obtains $\bar M=M$, in contrast with
what one finds, for instance, in the particular case of
spherically symmetric dilatonic black holes \cite{ch}
(see Appendix~\ref{large}).
This consideration supports the argument put forward in
Ref.~\cite{ch} that different conformal frames can be
distinguished by measuring the ADM mass.
\par
As in Section~\ref{Lag} we now analyze the effect of
conformal transformations on the boundary terms.
\subsection{String Frame and Einstein frame}
\label{Cse}
Adding a minimally-coupled scalar field does not change the
gravitational part of the action, therefore one finds
$E_{EH}[\gamma_{ij}]$ in the corresponding total Hamiltonian
$H'_{EF}$.
The ADM decomposition is again understood and it is known that
the conformal rescaling (\ref{Tgg}) is then a canonical
transformation \cite{garay} provided second time derivatives
of the three-metric are eliminated from the action.
One is then left with a surface term on $\partial\Sigma_t$
which yields $E_{EF}[\gamma_{ij}]=E_{EH}[\gamma_{ij}]$.
\par
The action $S'_{EF}$ then transforms to the SF according to
Eq.~(\ref{bSF1}), which we can now rewrite as
\be
S'_{SF}[\bar g_{\mu\nu},\phi]=S_{SF}[\bar g_{\mu\nu},\phi]
-\int_{t_1}^{t_2} dt\,E_{EH}[\bar g_{\mu\nu}]
\ ,
\ee
and one obtains
\be
E_{SF}[\bar g_{\mu\nu},\phi]=E_{EH}[\bar g_{\mu\nu}]
\ .
\label{E_SF}
\ee
This proves that the ADM mass can be computed by simply replacing
the EF metric with the SF metric inside Eq.~(\ref{E_EH}).
\par
In Appendix~\ref{large} we show that for dilatonic black holes,
since $\Omega=e^{\phi/2}$ is of the form in Eq.~(\ref{O_large}), one
measures a different ADM mass in the SF with respect to the EF
\cite{ch}.
\subsection{Non-minimally coupled scalar field}
\label{Cnmc}
Under the conformal transformation defined by Eqs.~(\ref{Tg})
and (\ref{Tphi}), the EF action with a minimally coupled scalar field
transforms according to Eq.~(\ref{tSB}), and yields
\be
\tilde S'[\tilde g_{\mu\nu},\tilde\phi]=
\tilde S[\tilde g_{\mu\nu},\tilde\phi]
-\int_{t_1}^{t_2} dt\,E_{EH}[\tilde g_{\mu\nu}]
\ .
\ee
Again, one finds that
\be
\tilde E[\tilde g_{\mu\nu},\tilde \phi]=
E_{EH}[\tilde g_{\mu\nu}]
\ .
\label{tE}
\ee
\par
In general one expects that $\phi\sim r^{-1}$, therefore
$\Omega_\xi$ does not contain terms of order $r^{-1}$ and
the ADM mass is not changed by the transformation
in Eqs.~(\ref{Tg}) and (\ref{Tphi}).
\section{Conclusions}
\setcounter{equation}{0}
\label{conc}
In this paper we have analyzed in detail the surface terms which arise
when performing certain conformal transformations that relate the
SF to the EF and the latter with a minimally coupled scalar field
to a non-minimally coupled one.
We showed that such transformations do not give rise to boundary
terms, except those which are needed to eliminate second derivatives.
\par
Boundary terms must be handled with particular care in the canonical
formalism, since they are related to the total energy of a system in
asymptotically flat space.
We have seen that the values taken by such terms generally depend
on the conformal frame and, in particular, the ADM mass of dilatonic
black holes in the EF is different from the one in the SF.
This fact is not surprising (see Appendix~\ref{large} and
Ref.~\cite{ch}), however, it conspires to question which frame is
physical \cite{faraoni,ch}.
\par
Finally, despite the fact that our analysis was purely classical,
one might consider the effect of surface terms in a quantum
context and argue that they can be related to different sectors of
the Hilbert space of states (see Ref.~\cite{c} for analogous
considerations).
We wish to return to this point in the future.
\appendix
\section{Conformally coupled case}
\setcounter{equation}{0}
\label{1/6}
The general relation (\ref{Tphi}) between $\phi$ and $\tilde\phi$ is
rather involved, however, the special case of conformal coupling
($\xi=1/6$) allows the simple expressions \cite{bekenstein}
\be
\tilde\phi=\tanh\phi
\ \ \ \ \ {\rm or} \ \ \ \ \
\tilde\phi={\rm cotanh\,}\phi
\ .
\ee
Correspondingly one has
\be
\Omega_{1/6}={1\over\cosh\phi}
\ \ \ \ \ {\rm or} \ \ \ \ \
\Omega_{1/6}=-{i\over\sinh\phi}
\ .
\label{O1/6}
\ee
\par
In asymptotically flat spaces, $\phi\simeq C\,r^{-1}$ for large $r$ and
only the first case in Eq.~(\ref{O1/6}) is acceptable, with
\be
\Omega_{1/6}\simeq 1+{C^2\over r^2}
\ .
\ee
Since there is no term of order $r^{-1}$, the ADM mass does not
depend on the conformal frame, as was already pointed out at the end
of Section~\ref{Cnmc}.
\section{Examples}
\setcounter{equation}{0}
In this Appendix we consider two cases with the purpose of showing
how surface terms in the Hamiltonian formalism behave.
In the first case (Section~\ref{frw}) we just have second time
derivatives of the three-metric which must be eliminated by adding
a surface term along the hypersurfaces of initial and final time;
in the second case (Section~\ref{large}), instead, we analyze surface
terms at large distance from a central source.
\subsection{Friedmann-Robertson-Walker space-time}
\label{frw}
For the case of a non-minimally coupled scalar field in
(spatially-flat) FRW space-time, the action (\ref{tS})
for $\tilde a=\tilde a(t)$ (scale factor of the Universe)
and $\tilde\phi=\tilde\phi(t)$ reduces to
\be
\tilde S=
{1\over 2}\,\int_{t_1}^{t_2}\,dt\,\tilde N\,\tilde a^3\,\left\{
{\dot{\tilde\phi}^2\over\tilde N^2}
+{6\over\tilde a^2}\,\left(1-\xi\,\tilde\phi^2\right)\,
\left[{\dot{\tilde a}^2\over\tilde N^2}
+{\tilde a\over\tilde N}\,{d\over dt}
\left({\dot{\tilde a}\over \tilde N}\right)\right]
\right\}
\ ,
\ee
in which the shift functions $\tilde N^i$ have been set to zero
and $\tilde N=\tilde N(t)$.
Upon getting rid of second time derivatives, according to the
prescription (\ref{pre}) with
\be
\sqrt{\tilde\gamma}\,\tilde K=
-3\,\tilde a^2\,{\dot{\tilde a}\over\tilde N}
\ ,
\ee
one obtains \cite{acg}
\be
\tilde S'=
{1\over 2}\,\int_{t_1}^{t_2}\,dt\,\tilde N\,\tilde a^3\,\left[
{\dot{\tilde\phi}^2\over\tilde N^2}
-{6\over\tilde a^2}\,\left(1-\xi\,\tilde\phi^2\right)\,
{\dot{\tilde a}^2\over\tilde N^2}
+12\,\xi\,{\tilde\phi\over\tilde a}\,
{\dot{\tilde a}\,\dot{\tilde\phi}\over \tilde N^2}
\right]
\ .
\ee
\par
The conformal transformation of the metric defined by Eq.~(\ref{Tg})
then becomes
\be
\tilde N=\Omega_\xi\,N
\ ,\ \ \ \ \
\tilde a=\Omega_\xi\,a
\ ,
\ee
and yields the action
\be
S'_{EF}=
{1\over 2}\,\int\,dt\,N\,a^3\,\left[
{\dot\phi^2\over N^2}
-{6\,\dot{a}^2\over a^2\,N^2}
\right]
\ ,
\label{FRW_EF}
\ee
in which no second time derivatives appear as well.
\par
The same form as in Eq.~(\ref{FRW_EF}) can be arrived at
from the SF action
\be
S_{SF}=
{1\over 2}\,\int\,dt\,\bar N\,\bar a^3\,e^{-\phi}\,\left[
{6\,\dot{\bar a}^2\over \bar a^2\,\bar N^2}
+{6\over \bar a\,\bar N}\,
{d\over dt}\left({\dot{\bar a}\over \bar N}\right)
-{\dot\phi^2\over 2\,\bar N^2}
\right]
\ ,
\ee
by first subtracting second time derivatives,
\be
S'_{SF}=
{1\over 2}\,\int\,dt\,\bar N\,\bar a^3\,e^{-\phi}\,\left[
{6\,\dot{\bar a}\,\dot\phi\over \bar a\,\bar N^2}
-{6\,\dot{\bar a}^2\over \bar a^2\,\bar N^2}
-{\dot\phi^2\over 2\,\bar N^2}
\right]
\ ,
\ee
and then defining
\be
\bar N=e^{\phi/2}\,N
\ ,\ \ \ \ \
\bar a=e^{\phi/2}\,a
\ .
\nonumber
\ee
\subsection{Black holes}
\label{large}
The classical example which is used to show the role played
by boundary terms in the canonical formalism is given by
the Schwarzschild metric, which is of the form given by
Eqs.~(\ref{g3}) and (\ref{N=f}) with
\be
N^2=1-{M\over 4\,\pi\,r}
\ .
\label{schw}
\ee
Any metric satisfying the conditions (\ref{asy}) approaches
the above form for $r\to\infty$ and one then finds
\be
E_{EH}&=&
-\int d\theta\,d\varphi\,\left(r^2\,N\,N'+2\,r\,N^2\right)_{r=R}
\nonumber \\
&\simeq&
\left[-{M\over 2}+\left(2\,M-16\,\pi\,r\right)\right]_{r=R\gg M}
\nonumber \\
&=&{3\over 2}\,M-16\,\pi\,R
\ ,
\ee
where $M$ is the ADM mass of the black hole.
The diverging term (for $R\to\infty$) in the r.h.s. of the above
equation can be eliminated by subtractiong the flat space
contribution \cite{gh}
\be
E_{EH}[g_{\mu\nu}]\to E_{EH}[g_{\mu\nu}]-
E_{EH}[\eta_{\mu\nu}]
&=&-\int d\theta\,d\varphi\,\left[\sqrt{\gamma_R}\,K_R
-r^2\,\sin\theta\,{4\over r}\right]_{r=R\gg M}
\nonumber \\
&=&
{3\over 2}\,M
\ .
\ee
\par
A less trivial example is given by the spherically symmetric charged
black hole \cite{gm,hs}, whose metric in the EF is given
by~\footnote{The case shown here corresponds to the choice
$a=1$ for the dilaton coupling (see Ref.~\cite{ch} for more details).}
\be
ds^2=-\left(1-{r_+\over r}\right)\,dt^2
+\left(1-{r_+\over r}\right)^{-1}\,dr^2
+r^2\,\left(1-{r_-\over r}\right)\,d\Omega^2
\ ,
\ee
where
\be
\begin{array}{l}
r_+=\strut\displaystyle{M\over 4\,\pi}
\\
\\
r_-=2\,\strut\displaystyle{Q^2\over r_+}
\ ,
\end{array}
\ee
and the dilaton field is
\be
e^{\phi}=1-{r_-\over r}
\ .
\ee
The ADM mass computed from the large $r$ expansion of the
metric is thus given by $M$ and coincides with the quantity
obtained from Eq.~(\ref{E_EH}),
\be
M={2\over 3}\,E_{EH}[g_{\mu\nu}]=4\,\pi\,r_+
\ ,
\ee
\par
Changing to the SF yields a different ADM mass, namely
\be
\bar M=M+{Q^2\over 2\,M}
\ ,
\ee
which is again obtained both from the large $r$ expansion of
\be
\bar g_{tt}=
-\left(1-{r_+\over r}\right)\,
\left(1-{r_-\over r}\right)
\ ,
\ee
and Eq.~(\ref{E_SF}),
\be
\bar M={2\over 3}\,E_{EH}[\bar g_{\mu\nu}]
=4\,\pi\,\left(r_++r_-\right)
\ .
\ee
\end{document}